\documentstyle[12pt]{article}
\input html.sty
\begin{document}
\title{MATTERS OF GRAVITY, The newsletter of the APS TIG on Gravitation}
\begin{center}
{ \Large {\bf MATTERS OF GRAVITY}}\\
\bigskip
\hrule
\medskip
{The newsletter of the Topical Group in Gravitation of the American Physical 
Society}\\
\medskip
{\bf Number 11 \hfill Spring 1998}
\end{center}
\begin{flushleft}

\tableofcontents
\bigskip
\hrule
\vfill
\section*{\noindent Editorial policy:}

The newsletter publishes articles in three broad categories, 

1. News about the topical group, normally contributed by officers of the 
group.

2. Research briefs, comments about new developments in research,
typically by an impartial observer. These articles are normally
by invitation, but suggestions for potential topics and authors are welcome
by the correspondents and the editor.

3. Conference reports, organizers are welcome to contact the editor or
correspondents, the reports are sometimes written by participants in the
conference in consultation with organizers.

Articles are expected to be less than two pages in length in all categories.

Matters of Gravity is not a peer-reviewed journal for the publication
of original research. We also do not publish full conference or meeting
announcements, although  we might consider publishing a brief notice 
with indication of a web page or other contact information.

\section*{\noindent  Editor\hfill}

\medskip
Jorge Pullin\\
\smallskip
Center for Gravitational Physics and Geometry\\
The Pennsylvania State University\\
University Park, PA 16802-6300\\
Fax: (814)863-9608\\
Phone (814)863-9597\\
Internet: 
\htmladdnormallink{\protect {\tt pullin@phys.psu.edu}}
{mailto:pullin@phys.psu.edu}\\
WWW: \htmladdnormallink{\protect {\tt http://www.phys.psu.edu/PULLIN}}
{http://www.phys.psu.edu/PULLIN}
\begin{rawhtml}
<P>
<BR><HR><P>
\end{rawhtml}
\end{flushleft}
\pagebreak
\section*{Editorial}

This newsletter includes for the first time an editorial policy. This
was formulated in consultation with the correspondents and in response
to an increased number of unsolicited research papers and conference
announcements.  I also want to apologize for the delay in publication
of this newsletter (it was due the first of the month). Some hackers
created a chat room in one of our workstations, and that interfered
with the production. As usual I wish to thank the contributors and
correspondents for making the newsletter possible. The newsletter
ended up being a bit too long for my taste, my apologies to those of
you who have insisted over time that I should keep it brief. I will
work harder on this in the future.

The next newsletter is due September 1st.  If everything goes well this
newsletter should be available in the gr-qc Los Alamos archives under
number gr-qc/9802017. To retrieve it send email to 
\htmladdnormallink{gr-qc@xxx.lanl.gov}{mailto:gr-qc@xxx.lanl.gov}
(or 
\htmladdnormallink{gr-qc@babbage.sissa.it}{mailto:gr-qc@babbage.sissa.it} 
in Europe) with Subject: get 9802017
(numbers 2-10 are also available in gr-qc). All issues are available in the
WWW:\\\htmladdnormallink{\protect {\tt
http://vishnu.nirvana.phys.psu.edu/mog.html}}
{http://vishnu.nirvana.phys.psu.edu/mog.html}\\ 
A hardcopy of the newsletter is
distributed free of charge to some members of the APS
Topical Group on Gravitation. It is considered a lack of etiquette to
ask me to mail you hard copies of the newsletter unless you have
exhausted all your resources to get your copy otherwise.

If you have comments/questions/complaints about the newsletter email
me. Have fun.
\bigbreak

\hfill Jorge Pullin\vspace{-0.8cm}
\section*{Correspondents}
\begin{itemize}
\item John Friedman and Kip Thorne: Relativistic Astrophysics,
\item Raymond Laflamme: Quantum Cosmology and Related Topics
\item Gary Horowitz: Interface with Mathematical High Energy Physics and
String Theory
\item Richard Isaacson: News from NSF
\item Richard Matzner: Numerical Relativity
\item Abhay Ashtekar and Ted Newman: Mathematical Relativity
\item Bernie Schutz: News From Europe
\item Lee Smolin: Quantum Gravity
\item Cliff Will: Confrontation of Theory with Experiment
\item Peter Bender: Space Experiments
\item Riley Newman: Laboratory Experiments
\item Warren Johnson: Resonant Mass Gravitational Wave Detectors
\item Stan Whitcomb: LIGO Project
\end{itemize}
\vfill
\pagebreak

\section*{\centerline {Topical group news}}
\addtocontents{toc}{\protect\smallskip}
\addtocontents{toc}{\bf News:}
\addtocontents{toc}{\protect\smallskip}
\addcontentsline{toc}{subsubsection}{\it  Topical group news, by Jim Isenberg}
\begin{center}
    Jim Isenberg, GTG secretary, University of Oregon\\
\htmladdnormallink{jim@newton.uoregon.edu}
{mailto:jim@newton.uoregon.edu}
\end{center}
\parindent=0pt
\parskip=5pt

{\bf Election News}

This year, as well as future years, our election for officers will  
take place in February. The nominating committee is currently putting  
together a slate. When they have done so, the slate will be e-mailed  
to everyone, and there will be an opportunity for everyone to comment  
and make recommendations. Shortly after that, the election will take  
place.

{\bf Officers}

Below,  I am including a list of the TGGrav officers last year and  
this year. Note that the officers change at the time of the April  
meeting. Abhay Ashtekar will take over as Chair. The newly elected  
officers will also begin their service at that time.

First Set of Officers (1996-1997)

Chair: Beverly K Berger

Chair Elect: Kip S. Thorne,
Vice Chair: Abhay Ashtekar,
Secretary Treasurer (1996-1999): Jim Isenberg .

Delegates (1996-1999): Frederick Raab, Leonard Parker, (1996-1998):
 David Shoemaker, James Bardeen, (1996-1997): Robert Wald, Lee Samuel Finn.

Nominating Committee: David Shoemaker (chair), Jorge Pullin, Peter  
Bender

Second Set of Officers (1997-1998)

Chair:  Kip S.Thorne 

Chair Elect: Abhay Ashtekar,
Vice Chair: Rainier Weiss,
Secretary/Treasurer (1996-1999): Jim Isenberg.

Delegates (1997-2000): Lee Samuel Finn, Mac Keiser,
(1996-1999): Frederick Raab, Leonard Parker,
(1996-1998): David Shoemaker, James Bardeen.

Nominating Committee: Fred Raab (chair), Jorge Pullin, Eric Poisson, 
                                       Jennie Traschen

{\bf April Meeting}

Elsewhere in MOG, the schedule of our sessions in the April meeting
is given. Further details can be found on the webpage of our Topical
Group.  Please come !

{\bf Prizes?}

There is some discussion of the possibility of setting up a prize (or  
even two), for work in gravitational physics. The restrictions set by  
the APS on such prizes are a bit rigid, but many of us think it is a  
good idea. We welcome suggestions and comments. Please send them to  
me (
\htmladdnormallink{jim@newton.uoregon.edu}
{mailto:jim@newton.uoregon.edu}).

{\bf Centenary  Speaker List}

The APS Centenary is coming up in 1999. To help celebrate, the APS is  
planning to set up a list of top notch speakers who will be called  
upon to give special colloquia and special public lectures around the  
country. If you wish to volunteer yourself or anyone else for  
inclusion on this list, please let me know.

\vfill
\pagebreak
\section*{\centerline {
The April Joint APS/AAPT Meeting: GTG Program}}
\addtocontents{toc}{\protect\smallskip}
\addcontentsline{toc}{subsubsection}{\it 
The April Joint APS/AAPT Meeting: GTG Program, by Abhay Asthekar}
\begin{center}
    Abhay Ashtekar, GTG program chair \\
\htmladdnormallink{ashtekar@phys.psu.edu}
{mailto:ashtekar@phys.psu.edu}
\end{center}
\parindent=0pt
\parskip=5pt

This year, the APS-AAPT will be held in Columbus Ohio from April
18 to 21, 1998. Our Topical Group will have the following activities
during this meeting: i) Three Invited Sessions; ii) Two Focus Sessions
with invited talks; iii) Three Sessions of Contributed Talks; iv) An
Executive Committee Meeting; and, v) Annual Business Meeting (the last
half hour of this meeting will be devoted to seeking input for the
decennial report, being prepared by Committee of Gravitational
Physics).

To facilitate your travel plans, we are enclosing the current program.
(However, please note that the APS has not finalized the schedule yet
and there may be some minor changes).  The titles of the invited talks
appeared on the GTG and MacCallum distribution lists and can also be
found on the GTG web site: 
\htmladdnormallink{\protect {\tt
http://vishnu.nirvana.phys.psu.edu/tig/}}
{http://vishnu.nirvana.phys.psu.edu/tig/}

\bigskip

{\it Saturday, April 18th}

$\bullet$ 11am - 2.00pm; Classical and Quantum Physics of Strong
Gravitational Physics (Focus Session 1); {\it Marolf, Rovelli, 
Isenberg, Moncrief}.

$\bullet$ 2:30 - 5:30pm; Extending the Frontiers of Gravitational
Physics (Invited Session); {\it Wolszczan, Friedman, Horowitz, Hartle}.
\medskip

{\it Sunday, April 19th}

$\bullet$ 8.30am - 10.30am; Gravitation Theory 1 (contributed Session)

$\bullet$ 10.45am -12:30pm; Executive Committee Meeting

$\bullet$ 11.00am - 2:00pm; Gravitation Theory 2 (Contributed Session)

$\bullet$  2.30pm - 5.30pm; Computation, General Relativity and 
Astrophysics \hfill\break\noindent
(GTG/DCOMP Joint Invited Session); {\it Matzner, Berger, Klein, Stone}.
\medskip

{\it Monday, April 20th}

$\bullet$ 8.30am - 11:00am; Gravitational Radiation: Confronting
Theory With Experiment (Focus Session 2); {\it Will, Price, Wiseman, 
Bender}.

$\bullet$ 2:30pm - 4:15pm; GTG Business Meeting
\medskip

{\it Tuesday, April 21st}

$\bullet$ 8:30am - 11:00am; Gravitational Experiments (Contributed 
Session) 

$\bullet$ 11:00am - 2:00pm; Precision Measurement Techniques
Applied to Fundamental Physics (Joint GTG/DAMOP Invited Session);
{\it Chu, Hall, Libbrecht, DeBra}.
\bigskip

In the last two APS meetings, the GTG sessions have been
lively. Through these sessions, we have initiated a healthy
interaction with the rest of the physics community. It is to our
advantage that the interaction continues to grow. We hope an even
greater number of GTG members will come to the next meeting.  See you
in Columbus!

\vfill
\pagebreak
\section*{\centerline {
Formation of the Gravitational-Wave} \\
\centerline{International Committee (GWIC)}}
\addtocontents{toc}{\protect\smallskip}
\addcontentsline{toc}{subsubsection}{\it 
Formation of the Gravitational-Wave International Committee,
by Sam Finn}
\begin{center}
    Lee Samuel Finn, Northwestern \\
\htmladdnormallink{lsf@holmes.astro.nwu.edu}
{mailto:lsf@holmes.astro.nwu.edu}
\end{center}
\parindent=0pt
\parskip=5pt

With several major new gravitational-wave detector projects nearing
completion in Europe, Japan and the United States, there is a need for
an organization where discussion regarding the international aspects
of experimental gravitational-wave physics can take place.

Recognizing this need, the directors of the five major interferometer
detector projects - ACIGA, GEO, LIGO, TAMA, and VIRGO - met on the two
days immediately preceding the Gravitational-Wave Data Analysis
Workshop to discuss the formation of such an organization. On
Wednesday, 12 November, with these aims in mind, they formed the
Gravitational-Wave International Committee, or GWIC.

GWIC's goals are to: 

* Promote international cooperation in all phases of construction and
    exploitation of gravitational-wave detectors;

* Coordinate and support long-range planning for new instrument
    proposals, or proposals for instrument upgrades;

* Promote the development of gravitational-wave detection as a
    astronomical tool, exploiting especially the potential for
    coincident detection of gravitational-waves and other fields
    (photons, cosmic-rays, neutrinos);

* Organize regular, world-inclusive meetings and workshops for the
    study of problems related to the development and exploitation of
    new or enhanced gravitational-wave detectors, and to foster
    research and development of new technology;

* Represent the gravitational-wave detection community
    internationally, acting as its advocate;

* Provide a forum for the laboratory directors to regularly meet,
    discuss, and plan jointly the operations and direction of their
    laboratories and experimental gravitational-wave physics
    generally.

GWIC's initial membership includes representatives of all the
interferometer detector projects (ACIGA, GEO, LIGO, TAMA, and VIRGO),
all the acoustic detector communities (ALLEGRO, AURIGA, EXPLORER,
GRAIL, NAUTILIS, and NIOBE), and the space-based detector community
(LISA). GWIC has a home-page on the web

\htmladdnormallink
{http://cithe502.cithep.caltech.edu/\~{}donna/GWIC/GWIC\_doc1.html}
{http://cithe502.cithep.caltech.edu/\~{}donna/GWIC/GWIC\_doc1.html},

where information on current and future activities can be found.

\vfill
\pagebreak
\section*{\centerline {
The 1997 Xanthopoulos Award}}
\addtocontents{toc}{\protect\smallskip}
\addcontentsline{toc}{subsubsection}{\it 
The 1997 Xanthopoulos Award, by Abhay Ashtekar}
\begin{center}
    Abhay Ashtekar, Penn State\\
\htmladdnormallink{ashtekar@phys.psu.edu}
{mailto:ashtekar@phys.psu.edu}
\end{center}
\parindent=0pt
\parskip=5pt

{\it Professor Matthew Choptuik} of the University of Texas at Austin
won this year's Basilis C. Xanthopoulos International Award in General
Relativity and Cosmology.

The Award was set up by the Foundation for Research and Technology
-Hellas in memory of Professor Xanthopoulos who was gunned
down (while giving a seminar) by a madman in 1990. It is given
tri-annually to a scientist, below 40 years of age, who has made
outstanding (preferably theoretical) contributions to gravitational
physics. The monetary value of the Award is approximately \$10,000.
The previous winners of the Award are Professors Demetrios
Christodoulou, Gary Horowitz and Carlo Rovelli. 

Over the past two years, the International Society for General
Relativity and Gravitation (GRG) and the Foundation for Research and
Technology -Hellas reached an agreement and from now on the prize be
presented by the President of the GRG Society during its tri-annual
conferences. Before each conference, the winner will be chosen by a
selection committee consisting of five to seven distinguished
scientists, each serving for two to three rounds. An advisory Board
will oversee the Award and ensure that the original intent of the
Award continues to be served.

Professor Choptuik received the Award during the last GRG conference
in Poona, India from the then President of the GRG Society, Professor
J\"urgen Ehlers. He was honored for his seminal contributions to
numerical relativity, in particular for the discovery of critical
phenomena associated with gravitational collapse.

\section*{\centerline {
We hear that...}}
\addtocontents{toc}{\protect\smallskip}
\addcontentsline{toc}{subsubsection}{\it 
We hear that..., by Jorge Pullin}
\begin{center}
    Jorge Pullin, Penn State\\
\htmladdnormallink{pullin@phys.psu.edu}
{pullin@phys.psu.edu}
\end{center}
\parindent=0pt
\parskip=5pt

Three members of our Topical Group were named Fellows of the American
Physical Society. I enclose email addresses so you can flood them with
congratulatory notes.

{\it Abhay Ashtekar}, 
\htmladdnormallink
{ashtekar@phys.psu.edu}
{mailto:ashtekar@phys.psu.edu}
nominated through the Topical Group on Gravitation, ``For 
his various contributions to classical and
           quantum gravitational physics, in particular the new
           canonical variables and the development of rigorous
           techniques for the quantization of gravity and other
           non-Abelian field theories.''

{\it Reinaldo Gleiser}, 
\htmladdnormallink
{gleiser@fis.uncor.edu}
{mailto:gleiser@fis.uncor.edu}
nominated through the Forum for International 
Physics ``For his role in the development of physics in C\'ordoba,
           and for his contributions to the application of exact
           solutions to Einstein equations and gravitational
           radiation theory.''
 
{\it Bill Hamilton}, 
\htmladdnormallink
{hamilton@phgrav.phys.lsu.edu}
{mailto:hamilton@phgrav.phys.lsu.edu}
nominated through the Topical Group on 
Instrumentation and Measurements ``For his wpioneering work and continuing 
leadership in
           developing gravitational-wave detectors, for
           back-action evading measurements of mechanical
           squeezed states, and for the development of techniques
           for magnetic shielding.''

\vfill
\pagebreak
\section*{\centerline {LIGO project update}}
\addtocontents{toc}{\protect\smallskip}
\addtocontents{toc}{\bf Research briefs:}
\addtocontents{toc}{\protect\smallskip}
\addtocontents{toc}{\protect\smallskip}
\addcontentsline{toc}{subsubsection}{\it 
LIGO project update, by David Shoemaker}
\begin{center}
    David Shoemaker, MIT\\
\htmladdnormallink{dhs@tristan.mit.edu}
{dhs@tristan.mit.edu}
\end{center}
\parindent=0pt
\parskip=5pt

Construction is nearing completion at the LIGO Hanford, Washington
site, and is in full swing at Livingston, Louisiana. The Hanford site
civil construction at the site (buildings, roads, power) is almost
complete (door bells are being installed), and the labs and technical
spaces are starting to fill.  At the Livingston site, the construction
of the buildings is largely finished and the forming of the concrete
covers for the beam tubes is well underway. 

Chicago Bridge and Iron, the company building the LIGO beam tubes
(which connect the vertex and ends of the two arms), has completed the
fabrication and installation of all 8 km of beam tube at the Hanford
site. Those tubes have been tested and are in the process of being
formally accepted. The fabrication equipment has been moved to a
facility near the Livingston site, and production is well underway.
Our contractor for the fabrication of the vacuum chambers and
associated equipment which will be in the located in the buildings,
Process Systems International, has installed many of the large
chambers and associated hardware for the Hanford site; testing is
starting. The vacuum chambers for the Livingston site are in
construction.

The sites are now home to permanent staff, and the Hanford Observatory
has now hosted to several LIGO-related meetings. It is extraordinarily
exhilarating to see the dreams of a gravitational wave observatory
turned into steel and concrete, and the scale of it all is
overwhelming.  The next meeting at the Hanford Observatory will be of
the LIGO Science Collaboration (or LSC), and is scheduled for March
12-13.

Fabrication of the LIGO Detector components is underway for parts of
the seismic isolation system, mirror suspensions, and optical
components. A large fraction of the critical test-mass mirrors have
been polished and coating will commence shortly. Testing of the first
article of one of the isolation system designs will take place early
in Spring 98, with production to shortly follow. Electronic designs
are being tested in prototype forms, and the first complete stabilized
Nd:YAG laser source is being assembled for delivery to the Hanford
site this summer.

A test of the phase-sensing system for LIGO is wrapping up in a
prototype interferometer at MIT. A record sensitivity of $ 1.5 \times
10^{-10}$ rad Hz$^{-1/2} $ has been demonstrated, using the basic
laser, suspension, and isolation technology planned for LIGO. This is
the last experiment in MIT's beloved Building 20 site, as the lab will
move this summer to a new location on the Campus; this enables a
reworked test interferometer which will help test second-generation
suspension and isolation concepts developed by the LSC.

Our schedule calls for shakedown of the interferometers starting in
mid-'99, and operation in 2001. Additional information about LIGO,
including our newsletter and information about the LSC, can be
accessed through our WWW home page at 
\htmladdnormallink{http://www.ligo.caltech.edu}{http://www.ligo.caltech.edu}.

\vfill
\pagebreak
\section*{\centerline 
{The Search for Frame-Dragging}\\\centerline{by Neutron Stars and Black Holes}}
\addtocontents{toc}{\protect\smallskip}
\addcontentsline{toc}{subsubsection}{\it 
Frame-dragging in NSs and BHs, by Sharon Morsink}
\begin{center}
Sharon Morsink, University of Wisconsin, Milwaukee\\
\htmladdnormallink{morsink@pauli.phys.uwm.edu}{morsink@pauli.phys.uwm.edu}
\end{center}
\parindent=0pt
\parskip=5pt

The dragging of inertial frames has been in the news lately with recent reports
that the Rossi X-ray Timing Explorer (RXTE) satellite has observed the  
gravitomagnetic precession of the inner edge of accretion disks around neutron
stars and black holes. If verified, this would be the first observation of a
strong field general relativistic effect. However, the result is far from 
conclusive with the present data. In this report, I'll give a short review
of the observations that have been made and describe some efforts to test
the hypothesis that frame-dragging has been seen. For a review of 
frame-dragging and efforts to measure the effect due to the Earth's motion,
see Cliff Will's article in MOG [1].

The truly exciting aspect of NASA's RXTE satellite is its ability to 
resolve  time variations in the x-ray spectrum occurring on time scales of 
order $0.1 ms$. Consider motion occurring at $r=6M$ outside of a 
$M=1.4M_\odot$  neutron star: test particles orbit with a  frequency of 
$\sim 1 kHz$ at this radius, corresponding to a time scale well within 
Rossi's resolution. Within the last two years, Rossi has discovered 
quasi-periodic oscillations (QPOs) occurring at repetition frequencies of 
$kHz$ order, suggesting that they are seeing phenomena  near  
neutron stars or black holes. A nice review of the $kHz$ QPO phenomenology
is given by van der Klis [2].
RXTE has seen $kHz$ QPOs from $14$ sources which are neutron stars in 
binaries. Their partners are difficult or impossible to observe, so the 
masses of these neutron stars aren't known. Typically, twin peaks in the 
Fourier analyzed x-ray spectrum are seen in these sources. (Take a look
at figure 4 of reference [2] for an example.) The peaks' 
frequencies (approximately $1 kHz$) drift with time, but their frequency 
separation stays constant.  A model, the sonic-point beat frequency model   
[3]  explains the twin peak phenomenon by identifying the higher
frequency peak with Keplerian motion of the accretion disk's inner edge. The 
peak separation is identified with the star's spin frequency. This leads to 
star rotation periods near $3 ms$. 
Some of these stars are occasional x-ray bursters and an analysis of the 
burst spectrum leads to a spin frequency which either agrees with the peak
separation or with twice the peak separation providing an independent check
of the model.

Suppose that the inner section of the accretion disk is 
tilted out of the star's equatorial plane. If this is the case, then 
the frame-dragging effect will 
cause the plane of the orbit to precess around the star, periodically 
obscuring the star. We would then expect to see a peak in the power spectrum
occurring at a frequency corresponding to the precession frequency. It was 
pointed out by Luigi Stella and Mario Vietri [4] that a peak with 
around the correct frequency appears in the spectrum. Moreover, they provide
a consistency check. As the inner edge of accretion disk changes location 
(due to radiation drag), the Keplerian frequency increases approximately 
as $2 \pi \nu_K = \sqrt{M/r^3} $ (remember that the star is 
rotating, so this is not exact). The Lense-Thirring precession varies as
$2 \pi \nu_{LT} = 2 J/r^3$, where $J$ is the star's angular momentum. 
Therefore, the peak which is to be identified with Lense-Thirring precession
should vary as the square of the Keplerian frequency peak. The data does 
show this rough trend. However, it is not this simple, since the star is not
spherical, and Newtonian gravity predicts a precession due to the star's 
quadrupole moment which subtracts from the frame-dragging precession 
frequency. Depending on the equation of state assumed for the neutron star, 
the quadrupole precession can range from a couple percent to half of the 
frame-dragging precession. Using a semi-Newtonian approximation, Stella
and Vietri found
that if the equation of state is very stiff, the data seemed to fit well. 
However, a more precise calculation, using general relativity [5],
 shows that the quadrupole (and higher multipole moments) become very 
important and greatly reduce the total precession. If the equation of state
is not overly stiff then the frame-dragging effect is dominant when the 
star is close to its maximum allowable mass. For typical equations 
of state, the total predicted precession frequency (including all effects)
is still only half of the peak's observed frequency. There is some 
possibility that
the factor of two could be explained by a geometric effect.  The system's 
geometry is essentially the same when the plane of the orbit has made a 
half period rotation, leading to a factor of two. However, this is still 
a bit speculative. In any case, astronomers are analyzing the RXTE data to 
find the observed variation of these peaks for a number of sources. If it 
should turn out that the dependence of the "precession" peak with the Keplerian
peak is correct, up to the factor of two, there may be some truth in the
model. It should be mentioned that a similar effect has been suggested in
the sources which correspond to alleged black holes [6],
but in these cases there are no twin peaks, so there is really no way 
to test the hypothesis.

There is a bit of a Catch $22$ [7] in the situation. Bardeen and
Petterson showed 
[8] that the combination of frame-dragging and viscosity produces
a torque which tends to align the disk with the star's equatorial plane,
so that Lense-Thirring precession won't occur. It is this effect which is
thought to keep the jets seen in active galactic nuclei aligned. Although
warped, precessing disks can occur, typically the inner part of the disk, 
up to $100M$ must be co-planar. If it is possible to find a physical 
mechanism which will cause a perturbation
to lift the inner edge of the disk, there will now be another force acting
on the inner edge of the disk. The precession frequencies computed assume
geodesic motion, i.e., that all other forces besides gravity are negligible. 
If precession occurs, the frequencies may change. Some work in this direction
indicates that this is the case [9,10], in fact 
reducing the possible frequencies by a large factor 
and/or  damping them strongly [10]. This is not to say that the peaks
observed can't be due to frame-dragging, but  it is difficult to  find a
physical mechanism which may cause a tilt without changing the frequencies.

In the meantime, we will have to wait for further analysis to learn whether 
there is a statistically significant correlation between the proposed 
precession peak and the Kepler peak. If so, it may be possible that 
frame-dragging has been observed near neutron stars.

{\bf References:}

[1] C. Will, The Search for Frame-Dragging, MOG No. 10, Fall 1997.\hfill\\ 
{[2]} M. Van der Klis, astro-ph/9710016.\\
{[3]} M.C. Miller, F.K. Lamb and D. Psaltis, astro-ph/9609157.\\
{[4]} L. Stella and M. Vietri, astro-ph/9709085.\\
{[5]} S.M. Morsink, L. Stella and M. Vietri, in preparation.\\
{[6]} W. Cui, S.N. Zhang and W. Chen, astro-ph/9710352.\\
{[7]} J. Heller, Catch 22, 1961.\\
{[8]] J.M. Bardeen and J.A. Petterson, ApJ 195, L65 (1975).\\
{[9]} J.R. Ipser, ApJ 458, 508 (1996); M.C. Miller, astro-ph/9801295.\\
{[10]} D. Markovic and F.K. Lamb, astro-ph/9801075\\

\vfill
\pagebreak
\section*{\centerline 
{Gamma-ray bursts: recent developments}}
\addtocontents{toc}{\protect\smallskip}
\addcontentsline{toc}{subsubsection}{\it 
Gamma-ray bursts: recent developments, by Peter Meszaros}
\begin{center}
Peter Meszaros, Penn State\\
\htmladdnormallink{nnp@astro.psu.edu}{nnp@astro.psu.edu}
\end{center}
\parindent=0pt
\parskip=5pt
{\it What are GRBs? }
Gamma-ray bursts (GRBs) are brief gamma-ray flashes detected with
space-based detectors in the range 0.1-100 MeV, with typical photon
fluxes of $0.01-100$ photons/cm$^2$/s and durations 0.1-1000
seconds. Their origin is clearly outside the solar system, and more
than 2000 events have been recorded so far.  Before there was any firm
evidence on the isotropy of classical gamma-ray bursts, the most
plausible interpretations involved magnetospheric events on neutron
stars (NS) within our Galaxy. However, the remarkable isotropy of
these events discovered within the last two years by the BATSE
experiment on the NASA Compton Gamma Ray Observatory (together with
the `flatter than Newtonian' counts) clearly shifts the odds
substantially in favor of a cosmological interpretation. Irrespective
of the distance (i.e., even in the galactic halo, but more so in
cosmological models), the energy density in a GRB event is so large
that an optically thick pair/photon fireball is expected to form,
which will expand carrying with itself some fraction of baryons
(e.g. Cavallo and Rees, 1978, Paczynski, 1986, Shemi and Piran,
1990). The main challenge in these models is not so much the ultimate
energy source (which may involve stellar collapse or binary compact
star merger) but rather how to turn the energy of a fraction of a
stellar rest mass into predominantly gamma rays with the right
non-thermal broken power law spectrum with the right temporal
behavior. The dissipative relativistic fireball model proposed by Rees
and Meszaros (1992, 1993, 1994; see also Narayan, Paczynski and Piran,
1992; Meszaros, Laguna and Rees, 1993, Meszaros, Rees and
Papathanassiou 1994, Katz, 1994; Sari, Narayan and Piran, 1996) is
largely successful in solving these problems, and is discussed in
several reviews, e.g.  Meszaros (1995, 1997).

{\it The Significance of GRB After-glows and Counterparts} 
The recent discovery (1997) of X-ray, optical and radio after-glows of
gamma-ray bursts (GRB) amounts to a major qualitative leap in the type
of independent observational hand-holds on these objects. Together with
existing gamma-ray signatures, these provide significantly more severe
constraints on possible models, and may indeed represent the light at
the end of the tunnel for understanding this long-standing puzzle of
astrophysics. The report of long wavelength observations of GRB 970228
over time scales of days to weeks at X-ray (X), and months at optical
(O) wavelengths (Costa etal, 1997) was the most dramatic recent
development in the field. In this and subsequent IAU circulars, it was
pointed out that the overall behavior of the long term radiation
agreed with theoretical expectations from the simplest relativistic
fireball afterglow models published in advance of the observations
(Meszaros \& Rees, 1997a). A number of theoretical papers were
stimulated by this and subsequent observations (e.g. Tavani, 1997;
Waxman, 1997a; Reichart, 1997; Wijers, etal, 1997, among others), and
interest has continued to grow as new observations provided apparently
controversial evidence for the distance scale, possible variability
and the candidate host (Sahu etal, 1997). New evidence was added when
the optical counterpart to the second discovered afterglow (GRB
970508) yielded a redshift lower limit placing it at a clearly
cosmological distance (Metzger etal, 1997), and this was strengthened
by the detection of a radio counterpart (Frail etal, 1997; Taylor
etal, 1997) as well as evidence for the constancy of the associated
diffuse source and continued power law decay of the point source
(Fruchter, etal, 1997). A third GRB afterglow (GRB971214) has also
been detected in X-rays and optical, and appears to follow the
canonical power law time decay (Heise, et al, 1997, and follwing IAU
circulars).  This new evidence reinforces the conclusions from
previous work on the isotropy of the burst distribution which
suggested a cosmological origin (e.g Fishman \& Meegan,
1995). Observational material on this is provided chiefly by a superb
data base (currently of over 1800 bursts in the 4B catalog) which
continues being accumulated by the BATSE instrument, complemented by
data from the OSSE and Comptel instruments on CGRO, as well as
Ulysses, KONUS and other experiments. At gamma-ray energies, much new
information has been collected and analyzed, relevant to the spatial
distribution, the time histories, possible repeatability, spectra, and
various types of classifications and correlations have been
investigated. At the same time, investigations of the physics of
fireball models of GRB have continued to probe the gamma-ray behavior
of these objects, as well as the after-glows. Much of the recent
theoretical work has concentrated on modeling the time structure
expected from internal and external shock models, multi-wavelength
spectra, the time evolution and the spectral-temporal correlations
(e.g. Papathanasiou \& Meszaros, 1996; Kobayashi, Sari and Piran 1997;
Waxman, 1997b; Katz \& Piran, 1997; Panaitescu \& Meszaros 1997a, 1997b;
Sari, Piran \& Narayan, 1997, etc.).

{\bf References:}\\
Cavallo, G. and Rees, M.J., 1978, M.N.R.A.S., 183, 359\\
Costa, E., 1997, IAU Circ. 6572; Nature, 387, 783 \\
Frail, D., et.al., 1997, Nature, in press \\
Fruchter, A. et al, 1997, IAU Circ. 6747 \\
Fishman, G. \& Meegan, C., 1995, A.R.A.A., 33, 415 \\
Heise, J, et al, 1997, IAU Circ. 6787; also 6788, 6789, 6791, 6795, etc\\
Katz, J., 1994, Ap.J., 422, 248\\
Katz, J. \& Piran, T., 1997, ApJ, in press\\
Kobayashi, T., Sari, R. \& Piran, T., 1997, ApJ, in press\\
Meszaros, P. and Rees, M.J., 1993, Ap.J., 405, 278; 1997a, ApJ, 476, 232 \\
Meszaros, P., Laguna, P. and Rees, M.J., 1993, Ap.J., 415, 181\\
Meszaros, P., Rees, M.J. and Papathanassiou, H., 1994, Ap.J., 432, 181\\
Meszaros, P., 1995, in 17th Texas Conf. Relativistic Astrophys, Bohringer, H.,
 etal, eds., (New York Acad. Sci., N.Y.), Ann.NY Ac.Sci, v.759, p.440 \\
Meszaros, P, 1997, in Proc. 4th Huntsville GRB Symposium 
 (AIP, in press, astro-ph/9711354)\\
Meszaros, P., Rees, M. J. \& Wijers, R. 1997, ApJ, in press 
(astro-ph/9709273)\\
Metzger, M et al., 1997, Nature, 387, 878 \\
Narayan, R., Paczynski, B. and Piran, T., 1992, Ap.J.(Letters), 395, L83\\
Paczynski, B., 1986, Ap.J.(Lett.), 308, L43\\
Panaitescu, A. \& Meszaros, P. 1997a,b ApJ, 492, ApJ(Lett)  
in press (astro-ph/9703187,9700284)\\
Papathanassiou, H and Meszaros, P, 1996, ApJ(Lett), 471, L91\\
Rees, M.J. and Meszaros, P., 1992, M.N.R.A.S., 258, 41P; 1994, Ap.J.
(Letters), 430, L93-L96\\
Reichart, D., 1997, ApJ, in press \\
Sahu, K., et al., 1997, Nature 387, 476 \\
Sari, R, Narayan, R and Piran, T, 1996, ApJ, 473, 204; 1997 
(astro-ph/9712005)\\
Shemi, A. and Piran, T., 1990, Ap.J.(Lett.), 365, L55\\
Tavani, M., 1997, ApJ(Lett), 483, L87 \\
Taylor, G.B., et al, 1997, Nature, in press \\
Vietri, M., 1997a, ApJ(Lett), 478, L9 \\
Waxman, E., 1997a, ApJ(Lett), 485, L5; 1997b, ApJ(Lett), in press (astro-ph/9709190)\\
Wijers, R., Rees, M.J. \& Meszaros, P., 1997, MNRAS, 288, L51 \\
\section*{\centerline 
{Moving Black Holes, Long-Lived Black Holes}\\
\centerline{ and Boundary Conditions:}
\centerline{Status of the Binary Black Hole Grand Challenge}}
\addtocontents{toc}{\protect\smallskip}
\addcontentsline{toc}{subsubsection}{\it 
Status of the Binary Black Hole Grand Challenge, by Richard Matzner}
\begin{center}
Richard Matzner, University of Texas at Austin\\
\htmladdnormallink{richard@ricci.ph.utexas.edu}
{richard@ricci.ph.utexas.edu}
\end{center}
\parindent=0pt
\parskip=5pt

The Binary Black Hole Grand Challenge has completed more than four
years of existence.  A large fraction of that time has been devoted to
developing a coherent infrastructure for assault on the two-black-hole
problem.  The Alliance approach involves a central Cauchy strong field
region, a boundary (matching) module, and an outer module
(perturbative, or strong-field characteristic) which carries the
radiation to infinity.  The interior module is an ADM
(Arnowitt-Deser-Misner [1] ) ``$\dot{g}, \dot{K}$'' code,
with fundamental variables the 3-metric $g_{ij}$ and extrinsic
curvature $K_{ij}$; the extrinsic curvature is the momentum of the
3-metric.  An extensive investigation has been made of the
Choquet-Bruhat/York [2]  version of a ``hyperbolic'' Cauchy
formulation, but the more traditional ADM form has the advantage of
more mature development.  Hence in 1996 the Alliance focused on the ADM
version.

Important infrastructure features include DAGH, which allows a
single-processor unigrid code to be distributed on a parallel machine,
and supports adaptive mesh refinement.  This system is now in use for
the very largest of our black-hole runs.  Another very important tool
is RNPL, which takes a high-level description of the physics and the
differencing scheme and generates C or FORTRAN code.  Another tool, on
the collaborative level, is SCIVIZ, which allows researchers to
collaborate to manipulate and visualize computational results.   
A new file format, SDF, has been developed, which overcomes efficiency and
size limitations of some other formats, for large parallel applications.

All of the Alliance codes are, and continue to be, demonstrated second
order convergent.  In all of our models, black holes are handled by
excising the domain inside the apparent horizon.  We have not yet begun
(we are about to begin) carrying out multiple black hole evolutions.
For single black holes (Schwarzschild or Kerr, and their strongly
perturbed forms), where we expect a stationary final state, we use the
analytic solution as an outer boundary condition.

{\it Black Holes}}
The characteristic module evolves the strong-field Einstein equations
in a characteristic formation which has a very rigid coordinate gauge,
and which therefore has a simpler equation set.  Unfortunately it
cannot be used alone for the binary black hole problem, since
gravitational focusing causes caustics in the rays generating the null
surfaces of the coordinatization.  Thus the basic approach of the
Alliance code is centered on a Cauchy strong field module.  However,
the characteristic code can handle {\it single} black hole
spacetimes.

The characteristic module  can exist in two forms based on either
ingoing or outgoing characteristic surfaces.  In its form based on
ingoing null hypersurfaces it shows unlimited long-term stability for
evolving single, isolated (distorted) black holes.  In these cases data
are set on an ingoing initial null hypersurface.  The inner edge of the
domain is set at a marginally trapped surface; no boundary condition is
needed there since it is inside the horizon.  The outer boundary is set
analytically to the black hole solution.  These problems evolve to the
stationary black hole form.  They have been evolved to times of $60,000
M$, at which time differences are on order of machine precision, the
operational definition of running forever.  In some cases the
coordinates are deliberately ``wobbled'' producing a time dependence in
the description at late times, but producing a stationary geometry
nonetheless [3].  In the 1970s and 1980s, the difficulty of
stably simulating even a single black hole in strictly spherical
symmetry (one spatial dimension) led to the formulation of ``the Holy
Grail of numerical relativity"  - requirements for a hypothetical
``code that simultaneously:

        $\bullet$ Avoids singularities

        $\bullet$ Handles black holes

        $\bullet$ Maintains high accuracy

        $\bullet$ Runs forever.''~[4]

It is clear that the characteristic code has achieved the grail in the
3-dimensional single-black-hole case, a dramatic improvement over the
state of the art only a few years ago.  However, goals recede, and from
the viewpoint of the Binary Black Hole Alliance, this is a step along
the way, important because it validates the stability and accuracy of
the characteristic code.

The interior code, the Cauchy code, has not yet shown the very
long-term stability of the characteristic code.  With fixed Dirichlet
boundaries, the code runs for a maximum [5]  of $100M$ for
isolated Schwarzschild data written in Kerr-Schild [6] form.  With
blended outer boundary conditions, the code has been evolved beyond
$500M$.  (The blended outer conditions are applied gradually by mixing
the computed results with the analytic ones over a shell of a few
computational zones' thickness; see also the discussion of this
technique for perturbative matching below.) In this case there is still
some influence from the outer boundary and there are additional modes
(small oscillations in the supposedly static solution) which are not
fully understood.  What is apparent is that inaccurate outer boundary
setting disturbs the code substantially (which is why the matching
algorithm is so important), but the inner edge of the domain, handled
with causal differencing (hence ``no boundary condition'') is well
behaved, and this free evolution shows (at worst) controllable
constraint drifting.

The Kerr-Schild data are represented by two fields on a background flat
space: a scalar function ($= M/r$ for Schwarzschild), and a null vector
(ingoing, unit for Schwarzschild).  Because of this very simple
structure, {\it boosting} these data is trivial, and we have used
such boosted initial data to start evolutions of black holes moving
across the computational domain.  So far as we know, only the Alliance
has achieved this.  The characteristic code has demonstrated a linearly
moving black hole [7].  However (because of the caustic
problem), the characteristic module cannot evolve a black hole moving
farther than one diameter.  The Cauchy module can do so, and has been
demonstrated to do so for $60M$ in time at $0.1c$, hence a translation
through $6M$ in distance [5].  The boundary conditions for
this moving case are analytical Dirichlet with no blending.  (Since we
know the analytic form for the boosted black hole as a function of
time, we compute new outer boundaries as a function of time for the
evolution.)

The black hole interior is excised in all our evolutions.  At the
resolutions we use (typically 60 to 100 grid zones in each direction),
there is room for only a few ($\sim 10-15$) points interior to the
black hole.  We find the best behavior when the hole is excised with a
buffer zone $\sim 5$ zones wide for both the moving and the stationary
evolutions.  Thus the excision of the interior occurs $\sim 5$ zones
inside the apparent horizon location.  This is probably relevant to the
fact that we do not lock the horizon coordinate location.  Rather, the
excision is based on the analytically expected coordinate location of
the horizon; and all our code crashes seem to be related to the
excised, un-evolved, region eventually extending beyond the horizon.
(This can happen because coordinate drift, which we do not attempt to
control, changes the coordinate location of the horizon, while our
excision domain has a {\it fixed} coordinate location.)  To our
knowledge only Daues [8] has demonstrated active horizon locking
in 3-dimensional black holes.  Daues achieved $\sim 140M$ non-moving
Schwarzschild black hole evolutions.  Implementing this tracking in the
Alliance code is a high priority and holds out the hope of even longer
evolutions.

\noindent{\bf Exterior Modules and Matching }

The {\it perturbative} exterior module is written in explicitly
Cauchy form.  The terms neglected in this perturbative module are
wave-wave interactions, while the background is explicitly modeled
(Kerr or Schwarzschild).  The matching to the Cauchy interior
{\it works} in this case; this matching has been demonstrated for
linear waves with very long evolution [9]; some more recent
results are at the Alliance web site (see below).  The matching is
accomplished in a way that correctly treats the outgoing nature of the
solution; in fact, the Sommerfeld condition is modified on its
right-hand side from $0$, to a contribution arising from the
perturbative outer evolution, so there is a strong similarity between
the perturbative and the characteristic boundary application.

In practice, the perturbative outer boundary match is handled in a
``thick'' shell.  At some radius $r_{E}$, the inner solution is
sampled.  These data are used for a perturbative evolution to a very
large radius $r_{outer} \simeq \infty$.  At a finite radius $r_{1} >
r_{E}$ begins the boundary region $r_{2} > r > r_{1}$.  The computed
inner solution is merged in this region with the value determined from
the exterior module. This provides a merged boundary condition on the
interior solution: that the Sommerfeld condition properly reflect the
terms describing backscatter, derived from the perturbative evolution.
{\it For the weak wave case this is a successful complete
expression of the inner-module/boundary/outer-module paradigm of the
Alliance philosophy.}

To match the {\it characteristic} module to the Cauchy inner module,
the outgoing characteristic form must be used.  (This {\it match}
has not yet been achieved.)  For outgoing radiation near the coordinate
outgoing null surfaces, the wave variables have slow variation, and the
system can be compactified so that infinity is a finite distance away
while still maintaining finite derivatives.  Hence, a characteristic
code can compute the whole exterior spacetime in a finite domain.  For
nonlinear scalar radiation [10], for spherical general
relativity [11], for cylindrically symmetric
relativity [12], and as we saw, for the weak field problem in full
3-d general relativity, the match has been carried out.  But so far a
stable match between the full 3-d strong-field Cauchy and
characteristic modules has not been achieved.  We are now attempting
such a match through blending, as in the successful perturbative case,
and there is hope that such an approach will work to match the Cauchy
and the characteristic codes.

Immediate future work involves setting data and beginning 2-hole
evolutions.  Because of the apparently better behavior of Kerr-Schild
formulated single holes, the initial data is being recomputed for this
case.  (These slices differ macroscopically from the ``standard''
conformally flat data that were solved completely prior to the
beginning of the Alliance [13].)  This work will proceed while
further runs for single holes continue.  The Cauchy module requires
standardization, validation against known behavior of distorted black
holes, and an explicit demonstration of its ability to evolve rotating
(Kerr) black holes.

Recent developments, including the points discussed here, are frequently 
posted to the Los Alamos preprint archive, and can also be found 
at the Alliance Web page:  

\htmladdnormallink{http://www.npac.syr.edu/projects/bh/}
{http://www.npac.syr.edu/projects/bh/}  \\
Select ``New developments.''

Richard Matzner is the Lead PI of the Binary Back Hole Grand Challenge
Alliance, NSF ASC/PHY 9318152 (arpa supplemented), which supported this
work.

{\bf References:}

[1] R. Arnowitt, S. Deser, C. W. Misner, in 
{\it Gravitation, an Introduction to Current Research},
L. Witten, ed. (Wiley, New York, 1962).\\
{[2]} Y. Choquet-Bruhat and J.W. York, ``Geometrical Well Posed 
Systems for the Einstein Equations,'' C.R. Acad. Sci. Paris, 
{\bf 321} 1089 (1995). \\
{[3]} The Binary Black Hole Grand Challenge Alliance,
``Stable characteristic evolution of generic \\
3-dimensional single-black-hole spacetimes", submitted to {\it Physical
Review Letters} (1998). gr-qc/9801069\\
{[4]} S.L. Shapiro, S.A. Teukolsky, in {\it Dynamical Spacetimes and
Numerical Relativity}, ed. J. Centrella (Cambridge UP, Cambridge, 1986)
p. 74.\\
{[5]}  The Binary Black Hole Grand Challenge Alliance
``Boosted three-dimensional black-hole evolutions with
singularity excision", {\it Physical Review Letters} (in press, 1998).
gr-qc/9711078\\
{[6]}  R.P. Kerr and A. Schild, ``Some Algebraically Degenerate 
Solutions of Einstein's Gravitational Field Equations,'' 
{\it Applications of Nonlinear Partial Differential Equations in 
Mathematical Physics}, Proc. of Symposia B Applied Math., Vol. XV11 
(1965).
 R.P. Kerr and A. Schild,``A New Class of Vacuum Solutions of the Einstein 
Field Equations,'' 
{\it Atti del Convegno Sulla Relativita Generale: Problemi 
Dell'Energia E Onde Gravitazionale}, G. Barbera, ed. (1965). \\
{[7]} R. Gomez, L. Lehner, R.L. Marsa, J. Winicour,
``Moving Black Holes in 3D", {\it The Physical Review} {\bf D56}, 6310 (1997).
 gr-qc/9710138 \\
{[8]} G. Daues, Ph.D. dissertation, Washington University, 
Saint Louis (1996). \\
{[9]} The Binary Black Hole Grand Challenge Alliance
``Gravitational wave extraction and outer boundary conditions by
perturbative matching", {\it Physical Review Letters} (in press, 1998).
gr-qc/9709082\\
{[10]} Nigel T. Bishop, Roberto Gomez, Paulo R. 
Holvorcem, Richard A. Matzner, Philippos Papadopoulos, and Jeffrey 
Winicour, ``Cauchy-characteristic matching: A new approach to 
radiation boundary conditions,'' {\it Physical Review Letters} 
{\bf 76} 4303 (1996).  \\
{[11]}  R. Gomez, R. Marsa and J. Winicour, ``Black hole 
excision with matching,'' {\it Physical Review D} {\bf 56}, 
(November 1997), gr-qc/9708002. \\
{[12]} C. Clarke, R. d'Inverno, and J. Vickers, {\it Physical 
Review D} {\bf 52}, 6863 (1995). \\
M. Dubal, R. d'Inverno, and C. Clarke,
{\it Physical Review D} {\bf 52}, 6868 (1995). \\
{[13]} G.B. Cook, M. W. Choptuik, M. R. Dubal, S. Klasky, 
Richard A. Matzner and S.R. Oliveira, ``Three-Dimensional Initial 
Data for the Collision of Two Black Holes,'' {\it Physical Review 
D} {\bf 47 } 1471-1490 (February 1993). 

\vfill
\pagebreak
\section*{\centerline 
{Quantum gravity at GR15}}
\addtocontents{toc}{\protect\smallskip}
\addtocontents{toc}{\bf Conference reports:}
\addtocontents{toc}{\protect\smallskip}
\addcontentsline{toc}{subsubsection}{\it 
Quantum gravity at GR15, by Don Marolf}
\begin{center}
Don Marolf, Syracuse University\\
\htmladdnormallink{marolf@suhep.phy.syr.edu}
{marolf@suhep.phy.syr.edu}
\end{center}
\parindent=0pt
\parskip=5pt

The appropriate starting point for this review is Carlo Rovelli's
plenary talk `quantum spacetime.'  Rovelli took upon himself the
unenviable task of commenting on the vast variety of approaches to and
aspects of the subject of quantum gravity.  In the first part of his
talk, we found Carlo wearing an unfamiliar hat -- that of an
experimental sociologist -- as he presented an inventory of the
preprints that appeared on hep-th and gr-qc during the first 10 months
of 1997.  With a total flux over 400 papers per month, roughly 1 in 4
addressed quantum gravity or related issues.  Breaking these up by
topic, he found 69 string papers per month, 26 loop gravity papers per
month, 8 on QFT in curved spacetime, 7 on lattice Quantum Gravity, and
perhaps 29 per month for all other aspects combined (with a given such
approach averaging no more than 5 papers per month).  In GR15, there
were four plenary talks (by Gibbons, Rovelli, Kozameh, and Zeilinger)
with quantum themes, as well as a large number of parallel sessions:
one afternoon of superstrings and supersymmetry, one afternoon of
quantum cosmology and conceptual issues, one afternoon of quantum
fields in curved spacetime and semiclassical issues, and two
afternoons of `quantum general relativity.'

Since Gary Gibbons gave a talk about M-theory (the theory formerly
known as `strings'), Rovelli spent most of his time discussing the
loop approach, though he did comment on QFT in curved spacetime,
dynamical triangulations, Regge Calculus, and other ideas.  I will
follow the results of his xxx experiment and address first string
issues, then loop issues, and finally other issues in quantum gravity.
Unfortunately, it will not be possible to discuss here more than a few
talks from the 5 afternoons of parallel sessions on quantum issues.

The plenary lecture by Gary Gibbons
gave a brief overview of what
has become known as M-theory; the lecture was quite well received.  Briefly, 
M-theory is a project arising out of string theory which is supposed
to be a more fundamental and, when complete, nonperturbative formulation
of quantum gravity.  Gibbons made an 
analogy between M-theory and a Northern European Medieval cathedral 
whose many parts, created by individual artisans, are works
of art on their own, but whose real beauty and structure are apparent
only when the cathedral is completed -- perhaps long after the deaths
of the earliest contributors.   M-theory is to be viewed as such a cathedral
under construction.  Some pieces are in place, and there are many
architects who share a common vision for what the cathedral will become.
However, the building process is far from complete, and Gibbons reminds
us that many cathedrals were completely redesigned as they were being built
so that, in the end, they bore little resemblance to the original conception.
Indeed, some designs were simply impossible to build.

Nevertheless, Gibbons emphasized the solidity of the of the foundation of 
M-theory (which rests on all of the successes of string theory, 
understandings of string duality, and the impressive calculations of
black hole entropy by Strominger, Vafa, etc.)
as well as the sweeping vision of the architects.  He also 
described the ``landscape and architecture of the partially completed
cathedral and of the surrounding countryside.''   His talk
focused on the relationship of M-theory with supergravity, and with various
BPS (aka supersymmetric) objects.  [The most commonly discussed 
supersymmetric objects are extremal black holes.]  Readers interested
in an introduction to this subject will surely enjoy the version of his
talk to be published in the conference proceedings.

The other major contribution to GR15 in the string/M-theory vein was
a review talk ``Strings and Semiclassical properties of black holes''
given by Gautam Mandal in the parallel session on superstring theory and
supergravity.  His review was necessarily short and condensed, but
fairly thorough.  The 1996 work on reproducing black hole thermodynamics
from string calculations was also nicely summarized in a talk given by
A. Dasgupta, who reported some new results on 
fermionic Hawking radiation in effective string models of black holes.
In addition, some observations about superstring inspired cosmological models
and the graceful exit problem for inflation were made by S. Bose.

Let us now return to Rovelli's discussion of loop quantum gravity.  He
stressed the fact that this approach is essentially non-perturbative so
that it could, in principle, provide a complete definition of the
theory.  However, this also means that it is difficult to compute
the kind of perturbative scattering results that are common in, for
example, string theory.  As major results, Rovelli described
the predicted quantization of areas and volumes, the
recent calculations of black-hole entropy by Ashtekar et. al., and the
fact that a set of constraints has been proposed which, if correct, 
could provide a complete non-perturbative definition of Quantum Gravity.

On the other hand, Rovelli also mentioned two difficulties:  one was
the lack of a general algorithm for computing physical results (such
as scattering phenomena) and the other was a concern over whether
the proposed constraints do indeed describe gravity or whether they
need to be modified or replaced in some way.  This concern was
largely based on the results of Lewandowski and Marolf showing 
that the algebra of the proposed constraints does not seem to match
the classical hypersurface deformation algebra (instead, it 
gives $[H(N), H(M)]=0$ for the commutator of two Hamiltonian constraints) 
and the corresponding work by Lewandowski, Marolf, Gambini, and Pullin.
This issue was a matter of some discussion both in the parallel session on
quantum general relativity and in informal discussion.  An overview
of the results was presented by J. Lewandowski, and comments
were made in the talks by T. Thiemann and J. Pullin.  As the subject
is still under consideration (and since I am a participant in this
discussion), I will summarize the comments only very briefly without
drawing particular conclusions:  Thiemann and Pullin 
each suggested a possible way to modify the loop
approach in order to improve the situation, while other comments were
made that, since the constraints themselves are not directly physical
observables, it is unclear exactly what physical problems the above algebra 
would cause.   Discussion continues, and should remain interesting.

A few words are now in order regarding other quantum aspects of the 
conference.  Kozameh's plenary talk on the null surface formulation of
GR was mostly classical, but described some recent results concerning
linearized quantum theory in this framework, in which the coordinates
of certain events become quantum operators.  He also expressed a hope
that this formulation will help to untangle deeper mysteries of quantum
gravity.

Without going into details, let me say that a high point of
the conference was the plenary talk by A. Zeilinger on precision experiments
using quantum correlations.  These ranged from classic EPR tests to
`quantum teleportation' -- all effects predicted by standard quantum
mechanics and verified in his laboratory.  A hope was expressed that, in
the near future, experimental techniques would be refined to the
extent that they could directly test Roger Penrose's ideas about 
the effects of gravity on quantum decoherence.  I would strongly
recommend a visit to Zeilinger's web site
at 

\htmladdnormallink{http://info.uibk.ac.at/c/c7/c704/qo/}
{http://info.uibk.ac.at/c/c7/c704/qo/}. 

Finally, a number of extremely interesting (non-string, non-loop)
papers were presented in the parallel sessions.  Unfortunately, there
is only space to mention a few of them here.  The talks by L. Ford, 
E. Flanagan, and S. Carlip seemed to be the most popular.  Very
Briefly, Ford reviewed the latest results on providing inequalities
that restrict the negative energy that states of a quantum field may
have in static spacetimes.   Flanagan discussed the (quantum) stability
of Cauchy horizons in 1+1 dimensions and described a necessary condition
for the horizon to be classically stable but quantum mechanically unstable.
Carlip discussed his recent paper in which he argues that, if a sum over
topologies is to be performed, the partition function for $3+1$ gravity
with negative
cosmological constant cannot converge, and that it is formally
analogous to a system with negative specific heat.  He also noted
that the formal role of the cosmological constant is similar to the
temperature of such a system, and this observation led him to speculate
that it might provide a mechanism for setting $\Lambda =0$.  The idea
is that, somehow, due to the negative `specific heat,' processes
that would normally increase $|\Lambda|$ would instead drive it to zero.

\vfill
\pagebreak
\section*{\centerline 
{An Experimentalist's Idiosyncratic Report on GR15}}
\addtocontents{toc}{\protect\smallskip}
\addcontentsline{toc}{subsubsection}{\it 
An Experimentalist's Idiosyncratic Report on GR15, by Peter Saulson}
\begin{center}
Peter Saulson, Syracuse University\\
\htmladdnormallink{saulson@suhep.phy.syr.edu}
{saulson@suhep.phy.syr.edu}
\end{center}
\parindent=0pt
\parskip=5pt

   It wasn't long ago that the knock on General Relativity
was that it was a theorist's playground, blissfully disconnected
from confrontation with experiment. If there was anyone who
was still unaware that things had changed, all she would have
had to do was attend the 15th International Conference on
General Relativity and Gravitation, held at IUCAA, Pune, India,
from December 16-21, 1997.

   In this reporter's opinion, GR15's suite of plenary talks
was the strongest and most varied of any international meeting
in years. Credit must be given to Ted Newman and the
Scientific Organizing Committee he chaired for fine choices
of topics and speakers. No fewer than ten invited talks were
devoted either fully or in large measure to observable phenomena.
Five of those were devoted in one way or another to gravitational
waves. Talks by Flanagan on the range of possible sources, by
Seidel and by Pullin on ways of calculating the gravitational
waveforms from the particularly interesting case of black hole
coalescences, and by Cerdonio and by Robertson on methods of
detection, together gave an unusually complete review of the
bustling state of this branch of activity. 

   But it was the other observationally flavored talks that gave
this meeting its most distinctive character. There were two talks
on aspects of gravitational lensing: the graceful review of the
astrophysical situation with which S.M. Chitre opened the
conference, and the whirlwind tour of the optics of caustics and
related subjects provided by Michael Berry. The latter 
included enough novel physics to send everyone away with something
new to think about; interferometer jocks will be investigating
the phase singularities near the waists of their Gaussian beams
with new interest. 

   There were two other unabashedly astrophysical plenary talks.
Malcolm Longair presented a personal overview of the state
of our knowledge of astrophysical cosmology, rooted in the
remarkable growth of observational knowledge that has occurred
in the past few years. All signs point to further dramatic
improvement in the situation, including further exploitation
of HST's capabilities and the expected detailed maps of
the Cosmic Background Radiation from the upcoming satellites
MAP and Planck. Ramesh Narayan performed the unlikely feat of
interesting a roomful of relativists in the subtleties of
energy transport in accretion disks, in the cause of achieving
something this reporter would have thought impossible a year
ago: demonstrating by conventional (X-ray) astronomy that
objects with event horizons inhabit known binary star systems.

   Special notice must be given to the talk farthest removed from 
the ordinary topics of a general relativity meeting, that of Anton
Zeilinger on experimental demonstrations of the spooky
non-locality of quantum mechanics. This is another subject that
has made dramatic progress in the past few years, 
most recently with the demonstration by Zeilinger's group of 
teleportation of a quantum state. The breakneck pace of progress
was made evident by Zeilinger's remark that the violations
of locality of the sort treated by Bell's Theorem were so
strong in their experiments (they were detected at the 100-sigma
level) that this phenomenon was used as a calibration.

   There was, as well, a rich set of contributed papers on
experimental topics. These were strongly dominated by talks
on gravitational wave detection, which in turn fell into two
classes: progress reports on the many interferometers now
under construction (LIGO, VIRGO, GEO600) or in the planning
stage (LISA, OMEGA), and reports brought back from the
trenches by the grizzled veterans already making gravitational 
wave observations. Among the latter, W. Hamilton and L. Iess
emphasized to the raw recruits the tricky issues posed by 
non-Gaussian noise statistics in, respectively, resonant-mass
detectors and spacecraft tracking experiments.

   General relativity has undoubtedly been enriched by its
new-found observational character. There is every reason to
hope that the next GR meeting, slated for summer 2001 in
Durban, South Africa, will be an occasion to share further
experimental progress in our subject.

\vfill
\pagebreak
\section*{\centerline 
{GR Classical}}
\addtocontents{toc}{\protect\smallskip}
\addcontentsline{toc}{subsubsection}{\it 
GR Classical, by John Friedman}
\begin{center}
John Friedman, University of Wisconsin, Milwaukee\\
\htmladdnormallink{friedman@uwm.edu}{friedman@uwm.edu}
\end{center}
\parindent=0pt
\parskip=4pt

This is by necessity a selective summary, based primarily on the
plenary talks, not because they encompass the most important results
reported at the meeting, but because of the author's limitations; among
other difficulties, overlapping parallel sessions mean that it is
impossible for one person to attend most of the relevant workshops.

{\it 1a.  Relativistic Astrophysics: Black holes}

Ramesh Narayan reviewed the current status of black-hole observation,
as well as advances in understanding accretion disks.  After recalling
the limit set by causality on the mass of spherical and rotating
neutron stars, he turned to a list of best candidates for stellar-size
black holes.  At present, the single best current candidate appears to
be V404 Cyg (Casares and Charles 1994), a low mass X-ray binary.  Among
X-ray binaries, V404 Cyg has the largest mass function known, $f(m) =
6.08\pm 0.06 M_\odot$, implying for the compact object a mass $12.3\pm
0.3 M_\odot$. The 9 best candidates include 7 LMXB's; and the narrow
error bars for several of these mean with near certainty that there is
a class of compact objects with mass well above the upper mass limit
for neutron stars (or any stars above nuclear density).   Two high-mass
X-ray binaries, Cyg X-1 and LMC X-3, made the list, but are no longer
the candidates to quote. (`High' and `low' refer to the mass of the
X-ray source's companion).

Vastly increased resolution in observations of the centers of galaxies
has, within the past five years, given us similarly compelling evidence
for super-massive black holes in the centers of 15-20 galaxies.  The
evidence suggests that nearly every large galaxy hosts a central black
hole.  Measured masses range from 2-3 million $M_\odot$ in the Milky
Way to 3 billion in M87.  Observations of NGC 4258 (Miyoshi et al 1995)
are an example of the extraordinary current resolution: $3.6\times 10^6
M_\odot$ lies within a diameter of 0.03 pc.

Narayan claimed a significant advance in our understanding
of accretion disks, with ``advection-dominated accretion flow'' models
giving striking agreement with observation for accretion below the 
Eddington limit on $\dot M$. When the density of accreting matter is 
low, infalling ions do not have enough collisions to transfer their
energy to the lighter electrons that could radiate it away.  Instead,
a substantial fraction of the infall energy is swallowed by the black
hole.  Narayan emphasizes that one indirectly sees the existence of 
a horizon in accreting black-hole systems: With a central star, 
simple energy bookkeeping implies a larger energy of infall than 
is observed in radiation.  Steady flow is consistent with observation
only if there is a horizon into which the energy can flow.

{\it 1b. Relativistic astrophysics: Numerical Relativity}

Ed Seidel presented an optimistic report on the Grand-Challenge project
to compute numerically the inspiral and coalescence of two black
holes.  Significant progress was reported in developing 3+1 codes that
use a grid that does not include black-hole interiors.  One
incorporates the lack of influence of a black-hole interior on the
exterior spacetime by causal differencing at the apparent horizon, and
3+1 evolutions of stationary and boosted black holes have run past
$t=1000 M$.  A first 3+1 evolution based on a foliation by null
surfaces and using the {\sl characteristic} initial value problem has
evolved Kerr and Schwarzschild spacetimes to $t = 20,000 M$, but the
code does not yet allow caustics.
Seidel did not have time to talk about corresponding work on the
analogous 3+1 evolution of neutron-star binaries, but substantial
progress by Oohara and Nakamura in the numerical relativity workshop.
(Others reporting advances on the neutron-star evolution problem were
Bonazzola et al and, for the grand-challenge group, Miller.)
A public-domain CACTUS code from the Grand Challenge group will soon be
available on a Web Server.

Jorge Pullin spoke about the recent development (with Price) of a
second-order perturbation theory of perturbations about a Schwarzschild
background (Tomita had previously developed a second-order formalism in
a Newman- Penrose framework).  Because a single horizon can surround
two black holes well before the individual apparent horizons meet,
perturbation theory can describe the coalescence of black holes with
unexpectedly high accuracy and for an unexpectedly large part of the
coalescence.  In fact, the second-order formalism accurately gave 
the {\sl phase} of waves emitted in the outgoing modes that dominate
black-hole ringdown.

Matt Choptuik won this year's Xanthopoulos Prize for his work on
Choptuik scaling and critical phenomena in black-hole formation, and
his talk summarized work in this area by a number of
people.  Critical behavior has recently been examined in a broader
class of settings.  For collapse in an Einstein-Yang Mills framework, one
again sees critical behavior (and discrete self-similarity) for
families of solutions that interpolate between no black hole and a
black hole of nonzero mass.  The critical exponent relating black-hole
mass near $M=0$ to a smooth parameter for the family is 0.20, clearly
different from the value(s) of 0.36 that were first seen in spherical
collapse of massless scalar fields and perfect fluids.  Collapse of
fields that have stationary solutions with nonzero mass show
mass gaps; this was suspected from, e.g., neutron stars, where
continuously adding mass pushes the star over the upper mass limit to a
black hole that first forms at about that limiting mass.  And a mass
gap is seen for massless quantum scalar fields in a QFTCST calculation
with back-reaction.

{\it 2.  Cosmology}

Malcolm Longair and Vladimir Lukash presented, with opposite
conclusions, summaries of recent cosmological observations.  Both
mentioned recent successes, many associated with the Hubble telescope,
in measuring with improved accuracy key cosmological parameters,
$\Omega_0\equiv  \rho_0/\rho_{\rm critical}$, $H_0$, $q_0$, $\Lambda$,
and the age $T_0$.  I'll pick out two things from Longair's
wide-ranging talk. First, the small dispersion of Type IA supernovae
(associated with the collapse of white dwarfs pushed over their upper
mass limit)  makes them ``a clear market leader'' as a standard candle
at large redshift.  Two 1997 supernovae tighten the evidence for an
open universe:

\begin{tabular}{|c|c|}\hline
Garnavich et al (1998) at $z=0.97$ 
and $z=0.83$ imply, & Perlmutter et al (1998)\\
If $\Omega_0 + \Omega_\Lambda = 1, \Omega_0<1$ at the 
$95\%$ confidence level.
& If $\Omega_0 + \Omega_\Lambda = 1, \Omega_0=0.6\pm 0.2$. \\
If $\Omega_\Lambda = 0$, $\Omega_0 = -0.1\pm 0.5$.&
If $\Omega_\Lambda = 0$, $\Omega_0 = 0.2\pm 0.4$.\\\hline
\end{tabular}

Second, the Hipparcos satellite's revision of the local distance scale 
means that stars are brighter (and hence burn faster) than had been thought.
The age of globular clusters is now $T_0 = (11.5\pm 1.3)\times 10^9$ years, 
consistent with the $H_0$ measurements.

S. M. Chitre presented a history of gravitational lensing with emphasis on 
its increasing role in cosmology.  Gravitational lenses now serve as 
tools for diagnosing the mass distribution of both luminous and dark 
matter and as giant telescopes that intensify objects at high redshifts.
  
{\it 3.  Classical Gravity}

Carlos Kozameh spoke about dynamics of null surfaces in GR. This is a
program pursued by Kozameh and collaborators over several years,
intended to reformulate the field equations as equations governing a
family of null surfaces.  The formulation uses a function $Z$
describing a sphere's worth of surfaces at each point of spacetime (or
at each point of phase space).  Recent applications of the formalism
involve specifying $Z$ in terms of radiative data at ${\cal I}$,
leading to an asymptotic approach to quantization that associates
operators with spacetime points.

\vfill
\pagebreak
\section*{\centerline 
{Bangalore gravitational wave meeting}}
\addtocontents{toc}{\protect\smallskip}
\addcontentsline{toc}{subsubsection}{\it 
Bangalore gravitational wave meeting, by Sharon Morsink}
\begin{center}
Sharon Morsink, University of Wisconsin, Milwaukee\\
\htmladdnormallink{morsink@pauli.phys.uwm.edu}
{morsink@pauli.phys.uwm.edu}
\end{center}
\parindent=0pt
\parskip=3pt
\vspace{-.3cm}
The Raman institute hosted a very pleasant and informative meeting on
December 11-12 1997, covering a number of topics of key importance in
gravitational wave astrophysics. The meeting's format consisted of ten
plenary sessions providing an overview of theoretical and
observational aspects of gravitational radiation.

The construction of LIGO, GEO and other interferometric gravitational wave
detectors has opened up the possibility of making astronomical observations 
using gravitational radiation. Harold Lueck presented a historical
overview of the development of the experimental techniques and technological
improvements which have made these detectors possible.  The signal-to-noise
ratio for these detectors will be quite low so sophisticated data analysis
methods will be needed. Addressing this problem, S. Dhurandhar discussed how
the matched filter method will be used to detect gravitational wave signals. 
He reviewed recent work in detecting signals artificially injected 
into sample background noise. B. Sathyaprakash provided us with an overview
of the types of sources which LIGO may be able to detect, including the
inspiral of compact binaries, pulsars and supernovae, to name a few. One of 
the possible sources are rapidly rotating neutron stars which become 
unstable due to the CFS mechanism. Nils Anderson provided an introduction 
to this instability and described his recent work which suggests that axial
perturbations may play as important a role as the polar modes. In the 
discussion of possible sources of gravitational radiation, it is usually 
assumed that general relativity is, in fact, the correct theory of gravity. 
Gilles Esposito-Farese reviewed the  extent to which general relativity 
has been tested. He stressed that although GR has been successfully tested 
in the weak field limit, there are scalar-tensor theories which agree with 
GR in the weak field but provide different strong field predictions, such 
as boson stars. 

The bulk of our knowledge of gravitational wave sources (e.g. compact binary
systems) comes from perturbation theory. The main problem of interest
 is to find the relation between the outgoing 
radiation and the matter and motion of the source. The remaining speakers 
discussed different aspects of this problem with reference to the inspiral 
of neutron star and black hole binaries. Blanchet discussed a method based
on matching of expansions in near, exterior and wave zones 
which he calls the Multipolar-post-Minkowskian approach. This approach 
allows one to calculate various non-linear non-local effects such as tail 
radiation, tails of tails and memory terms.
 Clifford Will
presented a different approach which he has dubbed DIRE (Direct Integration 
of the Relaxed Einstein Equations). DIRE addresses the problem of divergent 
integrals in the near and far zones and has been used to compute post-Newtonian
corrections to 3.5 order, agreeing with results discussed by Blanchet. 
Given that we know the radiation emitted by a source, can we predict the 
backreaction onto the system caused by this emission? This problem, known 
as radiation reaction, was discussed by Bala Iyer. He presented an approach
which assumes the validity of the principle of energy balance: the work
done by the reactive force is equal to the negative of the energy flux. 
It is important to verify that perturbation theory provides the correct results
for all problems which can be solved exactly. Misao Sasaki discussed black
hole binary 
inspiral in the limit that one of the black holes is much less massive than
the other. This approach has been used to show the validity of
 the  post-Newtonian expansion in this limit. The holy grail of numerical 
relativity is the exact computation of binary black hole mergers, and 
Ed Seidel reported on recent progress in this direction. In particular, he
focused on highly distorted isolated  black holes, and showed that the full
non-linear evolution agrees well with  the results of perturbation theory in
the regimes where perturbation theory should be valid. 
I would like to thank our hosts, Bala Iyer,  Joseph Samuel and all the 
students at the Raman Institute for organizing such an enjoyable and 
interesting conference!  

\vfill
\pagebreak
\section*{\centerline 
{Bangalore quantum gravity meeting}}
\addtocontents{toc}{\protect\smallskip}
\addcontentsline{toc}{subsubsection}{\it 
Bangalore quantum gravity meeting, by Domenico Giulini}
\begin{center}
Domenico Giulini, University of Z\"urich \\
\htmladdnormallink{
giulini@physik.unizh.ch}
{giulini@physik.unizh.ch}
\end{center}
\parindent=0pt
\parskip=5pt

On December 13.-14., just prior to GR15, the Raman Research Institute
at Bangalore in India hosted a discussion meeting on quantum general
relativity as part of its Golden Jubilee celebrations.
The plan was to have three talks each morning and one in the afternoon,
then followed by longer discussion sessions.
The beautiful setting of the institute, together with the un-forced
and smooth organization indeed created a perfect atmosphere for
inspiring discussions. The topics covered a fairly wide range,
from (2+1)-dimensional quantum gravity, loop gravity, lattice 
approaches and 3-dimensional topology to the quantum theory of 
black holes and, in particular, the issues associated with black 
hole entropy. Canonical approaches dominated the scene, but this 
was partly due to the unfortunate fact that Ashoke Sen had to 
cancel his talk on string calculations of black hole entropy.  

The first speaker was Steve Carlip who presented five main lessons 
that could so far be learned from (2+1)-dimensional gravity. He listed
numerous consistent ways for quantization and pointed out their partial 
inequivalences. For example, consistent quantizations with or without 
topology change exist, hence topology change is consistent with, but 
not required by, quantum gravity. Another striking lesson
concerns the euclidean path integral approach. In (2+1)-dimensions
it can be shown that the contribution from the many arbitrarily 
complicated interpolating topologies cannot be neglected 
(as is sometimes assumed). Once more it became clear that, 
despite all differences to (3+1)-dimensions, (2+1)-dimensional 
gravity is an important and useful test bed to study concepts and 
expectations in quantum gravity.

Carlo Rovelli gave a large scale survey on progress and problems 
in loop quantum gravity. Recent progress in physical predictions at 
the Planck scale mainly originate from calculations of spectra of 
operators (on the auxiliary Hilbert space of pure gravity) 
representing area and volume of two- and three-dimensional subsets. 
In absence of any matter degrees of freedom these subsets are 
mathematically specified in a non diffeomorphism invariant fashion. 
Progress on the mathematical side was also reported. The long standing 
problems concerning the lack of a scalar product, overcompleteness 
of the loop basis and the implementation of the reality conditions 
seem to be settled now. Anomaly free regularizations of the 
super-hamiltonian have been constructed, but there is still an 
ongoing debate as to its physical correctness, since it does not 
define a deformation of the classical constraint algebra and hence 
seems to reproduce the wrong classical limit. Rovelli ended by 
emphasizing the complementary strengths and weaknesses of loop 
quantum gravity and string theory. 

Renate Loll reported on the status of discrete approaches to
4-dimensional quantum gravity based on the Einstein action. 
She discussed results from Hamiltonian path-integral approaches 
with connection variables and dynamical triangulations. The common 
open problem is the absence of appropriate measures on the 
discretized configuration spaces. The choices explored so far 
seem too simple to lead to an interacting, diffeomorphism-invariant 
field theory.

There were two talks on topological issues in (3+1)-dimensional 
canonical gravity.
Domenico Giulini started with discussing the role and significance of 
three-dimensional topology in the classical and quantum theories.
One of the issues addressed was whether and how classical topology 
leaves its fingerprints in the quantum theory. In this context the 
mapping class groups of three-dimensional manifolds were argued to 
be the natural objects to look at, since they carry significant amounts 
of topological information and also enter the quantum theory through the 
reduction procedure. Giulini concluded by listing some general 
properties of 3d mapping class groups, like finite 
presentations, residual finiteness and semi-direct product structures.
Sumati Surya reported on some work using Mackey theory to find 
interesting representations of 3d mapping class groups and 
discussed their physical implications. Thinking of the 3-manifold 
as configuration of elementary `geons' (i.e. prime-manifolds), she 
showed and discussed the general absence of spin-statistics 
correlations at the kinematical level, and also the possibility 
of novel `cyclic' statistics types which she encountered with 
three RP-3 geons.

Two talks and an additional discussion session -- filling the gap 
that the cancellation of Ashoke Sen's talk left -- were devoted  
to black hole entropy. V.~Frolov's talk centered around the  
problem of universality of black hole entropy which, despite some  
impressive derivations, like e.g. by counting states of D-branes, 
is still an open one. He discussed the idea of entanglement entropy,  
some of its problems, and how they can be solved in some models
of induced gravity. He reported on recent work on such models showing 
that universality exists within a special class.
In Parthasarathi Majumdar's talk the different approaches to understand 
black hole entropy were compared. In particular, the string calculations 
and viewpoints now came to their right. A final discussion 
session, solely devoted to all kinds of questions relating to 
black hole entropy, marked the end of this most pleasant meeting.

\vfill
\pagebreak
\section*{\centerline 
{Cleveland Cosmology-Topology Workshop}}
\addtocontents{toc}{\protect\smallskip}
\addcontentsline{toc}{subsubsection}{\it 
Cleveland Cosmology-Topology Workshop, by Neil Cornish}
\begin{center}
Neil Cornish, Cambridge\\
\htmladdnormallink{N.J.Cornish@damtp.cam.ac.uk}
{N.J.Cornish@damtp.cam.ac.uk}
\end{center}
\parindent=0pt
\parskip=5pt

On a crisp fall weekend in Cleveland, an unlikely collection of
mathematicians and physicists met at Case Western Reserve University
to discuss the large scale topology of the universe. There was a
certain irony to the location, as the meeting was being held
just a short walk from where Michelson and Morely dispensed with the
ether one hundred years earlier, while the cosmologies being discussed
come with an absolute frame of reference. 

The main aim of the meeting was to foster closer ties between
geometers, cosmologists and theoretical physicists. Through this
exchange of ideas and expertise, we hoped to arrive at a better
understanding of the theoretical and observational characteristics
of multi-connected cosmologies.

The meeting ran to a workshop format with a small number of talks
providing a springboard for extensive and lively discussions. This
meant that all 30 participants did indeed participate, even though
only half the participants gave talks. One of the most active
participants was Bill Thurston, who got things rolling by taking us on
a tour of topology and geometry in dimensions 1 through 5. We learnt
how topology becomes more flexible with increasing dimensionality
while geometry becomes more rigid. The majority of Thurston's talk was
devoted to 3-manifolds, where both topology and geometry find their
optimal balance between flexibility and rigidity. Concepts such as the
prime decomposition of 3-manifolds were made accessible to the physics
audience by relating the underlying ball-gluing construction to
wormholes. Thurston emphasised that most 3-manifolds are
hyperbolic. Picking up on this lead, David Spergel reviewed the
mounting observational evidence that we live in a sub-critical
universe with hyperbolic spatial sections. Gary Gibbons took us back
to the quantum gravity epoch and considered how the universe might
arrive at a non-trivial topology. The mathematicians were introduced
to the Euclidean path integral approach and semi-classical real
tunnelling geometries. Steve Carlip continued in a similar vein, but
argued that the density of topologies might dominate the gravitational
action in the path integral. Later in the meeting John Freedman
discussed multi-connected spacetimes from a Lorentzian quantum gravity
perspective and Bai-Lok Hu described the Casimir and other finite size
effects. Closely related to this was Jean-Philippe Uzan's description
of the obstructions to forming topological defects, such as cosmic
strings, in universes with non-trivial topology.

The majority of the workshop was devoted to observational searches for
topology. The meeting organizer, Glenn Starkman, introduced this topic
with a historical review of global topology in cosmology, starting in
1917 with de Sitter's $RP^3$ variant to Einstein's static universe and
moving to the present. Since the best window on topology is provided
by the cosmic microwave background radiation, David Spergel provided a
review of CBMR physics and observations. We also heard how the MAP and
Planck satellites will transform our view of the CMBR early next
century. Speaking for the Toronto group, Turan Souradeep told us about
their efforts to model the CMBR power spectrum in multi-connected
hyperbolic universes. On the same topic, Janna Levin entertained us
with her quirky description of the work done by the Berkeley group,
showing how cusped manifolds lead to flat spots in the CMBR. Still on
this theme, I outlined the work Neil Turok and I have done to develop
a simple numerical method for finding the eigenmodes of arbitrary
compact manifolds. Moving to more direct detection methods, I
explained how Spergel, Starkman and I hope to use the MAP satellite to
search for topologically matched circles in the CMBR. Searching a lot
closer to home, Boud Roukema showed how matched quasar groupings could
be used to test for non-trivial topology. This method should be quite useful
when the Sloan and Quest digital sky surveys deliver millions of new
quasars positions. As a fitting testament to the cross-disciplinary
nature of the workshop, one of the most intriguing observational
prospects was described by the topologist Jeff Weeks. Using his {\em
SnapPea} computer program, Weeks showed how the size and position of
just a few matched circle pairs could be used to completely
reconstruct the topology of the universe. A similar procedure can
also be used with the quasar groupings.

All the talks stimulated enthusiastic discussion that brought in
the other participants, including the topologist Colin Adams, Rob
Meyerhoff, John Ratcliffe and Bill Goldman
and physicists Ted Jacobson, Tanmay Vachaspati and Rich Gott. By the
end of the workshop the topologists were arguing about the Sachs-Wolfe
effect and the physicists were arguing about Dehn surgery on cusps.
All the participants agreed that it was one of the best meetings they
had been to in years.
If you want to hear more about what went on at the meeting, keep an
eye out for the workshop proceedings that will be appearing as a
special issue of {\em Classical and Quantum Gravity}. Or you could do
the nineties thing and visit the workshop website at
\htmladdnormallink{http//theory5.phys.cwru.edu}{http//theory5.phys.cwru.edu}.
There you will find contributed talks, a copy of the Cleveland Plain
Dealer article covering the workshop (complete with a picture of Glenn
Starkman eating a hyperbolic potato chip) and perhaps an audio file of
the radio coverage by the CBC radio program ``Quips and Quarks'',
produced by Dan Falk.

\vfill
\pagebreak
\section*{\centerline 
{Quantum Gravity in the Southern Cone II}}
\addtocontents{toc}{\protect\smallskip}
\addcontentsline{toc}{subsubsection}{\it 
Quantum Gravity in the Southern Cone II, by Carmen Nu\~nez}
\begin{center}
Carmen Nu\~nez, IAFE, Buenos Aires\\
\htmladdnormallink{carmen@iafe.uba.ar}{carmen@iafe.uba.ar}
\end{center}
\parindent=0pt
\parskip=5pt
The second edition of the Quantum Gravity in the Southern Cone workshop
was held at the Centro At\'{o}mico Bariloche, Argentina on January 7-10,
1998. It brought together 50 researchers from South America as well as
experts from the northern hemisphere, working on different aspects of
quantum gravity and related topics. The plenary lectures enabled the
participants to obtain a global picture of the status of the field in the
various approaches. The meeting was further enriched by poster sessions.
The following list summarizes the topics covered by the lecturers.

{\it Canonical Quantum Gravity}

{\it J.Pullin} overviewed of the attempts to apply the rules of canonical
quantization to GR. He stressed the important role played by spin networks
and by Thiemann's Hamiltonian. The problems presented by this Hamiltonian
(it commutes on non diffeomorfism invariant states) were addressed by 
{\it R. Gambini,} who reported on progress made to solve them.
Midi-superspace models of canonical quantum gravity were considered by 
{\it C. Torre} who indicated that one can satisfactorily quantize quantum
parameterized field theories on a two-dimensional spacetime, but that the
quantization of such theories in higher dimensions is still an open problem.
A framework for modelling quantum gravitational collapse was discussed by 
{\it K. Kuchar }who considered the canonical dynamics of matter shells.
Both the dynamics of the shell and of the surrounding spacetime were shown
to follow from a single variational principle.
By formulating GR as a theory of surfaces, {\it C. Kozameh} showed how to
construct a quantum spacetime using only Scri equipped with free functions
as the kinematical structure

{\it String theory and higher dimensional objects}

{\it J. Maldacena} discussed the large N limit of certain field theories
and its relation to gravity. In a similar context, {\it A. Schwimmer}
referred to N=1 (Seiberg) duality in field theory and its realization
through branes.
Phenomenological aspects of string theory were covered by {\it G.
Aldazabal} who discussed non-perturbative orbifold vacua.
Branes in supergravities, string theory and M theory were
discussed by {\it M. Cederwall}. {\it M. Henneaux} referred to dyons,
charge quantization and electric-magnetic duality for $p$-form theories in $%
2(p+1)$ spacetime dimensions with arbitrary gauge invariant
self-interactions. {\it B. Carter} discussed the geometry of non null $p$
surfaces embedded in $n$ dimensions. Electric 2 branes were presented by 
{\it R. Aros}. {\it J. Zanelli} reviewed Chern-Simons supergravity in $(2n-1)$
dimensions, showing that they contain non trivial dynamics leading to
interesting classical solutions such as black holes, solitons, membranes,
etc.

{\it Black hole physics, semiclassical theories and cosmology}

In the context of a two dimensional exactly solvable model, {\it
J. Russo} outlined the construction of an S-matrix and showed that
black holes will radiate out an energy of Planck order, stabilizing
after a transitory period. A similar picture appears in 3+1 Einstein
gravity with spherical symmetry. {\it R. Bousso} discussed the
evaporation of Schwarzschild-De Sitter black holes including the
one-loop effective action.  {\it B.Hu} addressed the problem of
fluctuations and backreaction in semiclassical cosmology and black
holes by presenting a complete history of the subject and conjectured
that a stochastic description in terms of Einstein-Langevin equation
becomes relevant at the Planck scale.  Semiclassical theories were
also considered by {\it C. Molina-Paris} and {\it S. Ramsey}. {\it
H. Rubinstein }reviewed the status of the big bang standard model and
the latest data available from observations.

\vfill
\pagebreak
\section*{\centerline 
{Baltimore AMS meeting}}
\addtocontents{toc}{\protect\smallskip}
\addcontentsline{toc}{subsubsection}{\it 
Baltimore AMS meeting, by Kirill Krasnov}
\begin{center}
Kirill Krasnov, Penn State\\
\htmladdnormallink{krasnov@phys.psu.edu}
{krasnov@phys.psu.edu}
\end{center}
\parindent=0pt
\parskip=5pt

At the Baltimore meeting of the American Mathematical Society, there
was a special session on quantum gravity and low-dimensional topology.
The session was organized by J. Baez (University of California,
Riverside) and S. Sawin (Fairfield University). The session lasted two
days: January 7 and the morning of January 8. This and the other
sessions of the AMS meeting took place at Baltimore Convention Center,
located in the tourist attraction center of Baltimore called Inner
Harbor.

On the first day, {\it Louis H. Kauffman} spoke about ``Discrete
Physics'', {\it Abhay Ashtekar} on ``Quantum Theory of Riemannian
Geometry'' and {\it Jim Stasheff}, gave ``A Survey of Cohomological
Physics''.  {\it Roger Picken} spoke about ``Kontsevich Integrals,
Knot Invariants and TQFT'', {\it Lee Smolin} discussed how to get
``Perturbative Strings from Perturbations of Evolving Spin Networks''
{\it Alexander A. Voronov} presented ``The Homotopy Algebraic
Structure of Topological Gravity'' and {\it John W. Barrett} talked on
``Quantum Gravity: Path Integrals and state Sums''.  {\it Louis Crane}
presented ``A State Sum Formulation for Quantum General Relativity''
and {\it Seth A. Major} discussed his work with {\it Roumen Borissov}
on ``Q-Deformed Loop Representation for Quantum Gravity: Structure and
Open Problems''.  {\it Kirill V. Krasnov} discussed ``Spin Networks,
Chern-Simons Theory and Black Holes'' and {\it Donald M. Marolf}
presented his work with {\it Jerzy Lewandowski} ``Loop Constraints: A
Habitat and their Algebra''

In the evening of that day {\it Edward Witten} delivered his 
Josiah Willard Gibbs Lecture on {\it M Theory} to 
more than a thousand mathematicians gathered at the Ballroom of
the Convention Center.

The session continued in the morning of the next day with {\it David
N. Yetter} talking about a ``Grist for a 4-D State-Sum Mill: Examples
of Monoidal Bicategories'', {\it Carlo Rovelli} presenting ``From Loop
Quantum Gravity to a Sum over Surfaces'' and {\it Dana S. Fine}
discussing ``Path Integrals Linking Chern-Simons and WZW Partition
Functions''. {\it Laurel T. Langford} spoke on 
``2-Tangles as a Free Braided Monoidal 2-Category with Duals''
and {\it Fotini G. Markopoulou} discussed 
``Quantum Space and Causality'', {\it Takashi Kimura}
presented his work with {\it Alexandre Kabanov} on 
``Tautological Classes and Cohomological Field Theories in Genus One''
and {\it Doug Bullock} his work with  {\it Charles Frohman} and 
{\it Joanna Kania-Bartoszynska} on
``Lattice Gauge Field Theory and Deformation Quantization''. Finally,
{\it Steven J. Carlip} discussed 
``Einstein Manifolds, Spacetime Foam, and the 
          Cosmological Constant''.

The session was a very interesting mixture of mathematics and
physics: talks on Topological Quantum Field Theory and
Quantum Gravity usually followed each other. In fact,
it was quite surprising to see the growing influence of the two 
fields on one another: many of 
the talks on the Quantum Gravity side were devoted
to the application of the ideas and techniques from 
TQFT's to gravity and some talks given by mathematicians
were on issues that used to be of interest only to physicists.
Belonging to this last category were the two very exciting 
talks by J. Barrett and L. Crane on their
new state sum formulation of quantum general relativity.

After the session was finished, some of us gathered for an informal
discussion guided by J. Baez, L. Crane, C. Rovelli and L. Smolin.
The discussion was devoted to some aspects of the path integral
formulation of quantum gravity.

\end{document}